\documentclass[sigconf,natbib=false]{acmart}
\usepackage[utf8]{inputenc}
\usepackage{newunicodechar}
\newunicodechar{́}{\'}

\usepackage[table]{colortbl}
\usepackage{algorithm}
\usepackage{algpseudocode}
\usepackage{enumitem}

\copyrightyear{2024}
\acmYear{2024}
\setcopyright{rightsretained}
\acmConference[MMSys '24]{ACM Multimedia Systems Conference 2024}{April 15--18, 2024}{Bari, Italy}
\acmBooktitle{ACM Multimedia Systems Conference 2024 (MMSys '24), April 15--18, 2024, Bari, Italy}
\acmDOI{10.1145/3625468.3647613}
\acmISBN{979-8-4007-0412-3/24/04}

\RequirePackage[
  datamodel=acmdatamodel,
  style=acmnumeric,
  sorting=none,
  ]{biblatex}

\begin{document}

\title[DIGITWISE]{DIGITWISE: Digital Twin-based Modeling of Adaptive Video Streaming Engagement}

\author{Emanuele Artioli}
\email{emanuele.artioli@aau.at}
\orcid{1234-5678-9012}
\affiliation{
  \institution{Alpen-Adria-Universitaet}
  \streetaddress{Universitaetstrasse 3}
  \city{Klagenfurt}
  \state{Kaernten}
  \country{Austria}
  \postcode{9020}
}
\author{Farzad Tashtarian}
\email{farzad.tashtarian@aau.at}
\orcid{1234-5678-9012}
\affiliation{
  \institution{Alpen-Adria-Universitaet}
  \streetaddress{Universitaetstrasse 3}
  \city{Klagenfurt}
  \state{Kaernten}
  \country{Austria}
  \postcode{9020}
}
\author{Christian Timmerer}
\email{christian.timmerer@aau.at}
\orcid{1234-5678-9012}
\affiliation{
  \institution{Alpen-Adria-Universitaet}
  \streetaddress{Universitaetstrasse 3}
  \city{Klagenfurt}
  \state{Kaernten}
  \country{Austria}
  \postcode{9020}
}
\renewcommand{\shortauthors}{Artioli et al.}

\begin{abstract}
  As the popularity of video streaming entertainment continues to grow, understanding how users engage with the content and react to its changes becomes a critical success factor for every stakeholder. User engagement, i.e., the percentage of video the user watches before quitting, is central to customer loyalty, content personalization, ad relevance, and A/B testing.
  This paper presents DIGITWISE, a digital twin-based approach for modeling adaptive video streaming engagement. Traditional adaptive bitrate (ABR) algorithms assume that all users react similarly to video streaming artifacts and network issues, neglecting individual user sensitivities. DIGITWISE leverages the concept of a digital twin, a digital replica of a physical entity, to model user engagement based on past viewing sessions.
  The digital twin receives input about streaming events and utilizes supervised machine learning to predict user engagement for a given session. The system model consists of a data processing pipeline, machine learning models acting as digital twins, and a unified model to predict engagement. DIGITWISE employs the XGBoost model in both digital twins and unified models. 
  The proposed architecture demonstrates the importance of personal user sensitivities, reducing user engagement prediction error by up to 5.8\% compared to non-user-aware models. 
  Furthermore, DIGITWISE can optimize content provisioning and delivery by identifying the features that maximize engagement, providing an average engagement increase of up to 8.6 \%.
\end{abstract}



\received{20 September 2023}
\received[revised]{12 March 2024}
\received[accepted]{5 June 2024}

\maketitle

\section{Introduction}

\begin{figure}
    \centering
    \includegraphics[width=\linewidth]{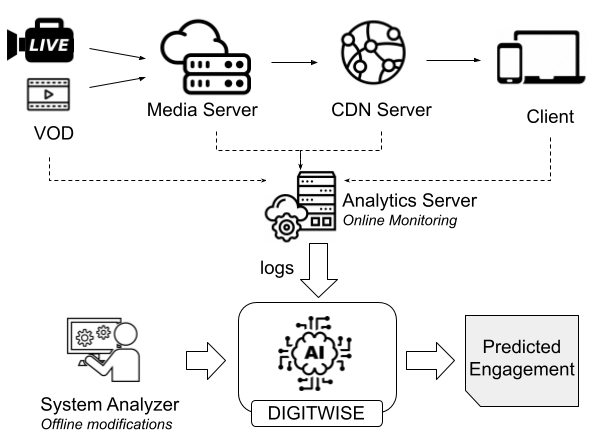}
    \caption{Conceptual architecture of video streaming with DIGITWISE enhancements.}
    \Description[architecture]{Conceptual architecture of video streaming with DIGITWISE enhancements}
    \label{fig:system-overview}
\end{figure}

Ever since video streaming emerged as one of the main applications of the Internet, with a current share of network traffic of around 65\%~\cite{Sandvine}, researchers have been looking for the optimal measure of viewer satisfaction.
HTTP Adaptive Streaming (HAS), is the leading approach for video streaming and involves a media server breaking down a video into segments and encoding each segment in various versions distinguished by bitrate and resolution settings~\cite{bentaleb2018survey}. 
These versions are then organized into a bitrate ladder, sometimes called an encoding ladder, which is essentially a structured list of pairs representing bitrate and resolution. 
Each client/user, also known as a player, requests the server for a manifest file, containing information about the bitrate ladder and other related data.
The delivery of the video occurs segment by segment: when a client requests the next segment, it selects a representation from the bitrate ladder based on an adaptive bitrate (ABR) algorithm. 
This selection aims to adapt to changing network conditions while balancing conflicting performance goals. 
For instance, opting for a higher bitrate can enhance video quality but may lead to buffering if the segment does not arrive in time for playback. 

\textbf{User Engagement.}
The knowledge of how and how much users are impacted by streaming events, such as rebuffering or visual artifacts, i.e., knowledge of user sensitivities, would allow video streaming providers to make better decisions in encoding and delivery. 
This in turn would save resources and increase viewer satisfaction and view time, which strongly correlates with revenue \cite{balachandran2013developing}.
Interestingly, sensitivities are rarely considered in the design of ABR algorithms, implicitly assuming that all users react the same way to the same streaming experience. 
Research shows this assumption is imprecise and that user sensitivity to artifacts, design choices, and network limitations is fundamental to their unique behavior~\cite{rodriguez2014content}. 
However, the discovery and modeling of user sensitivities remain challenging.
Initially, user engagement was modeled as uniform and solely dependent on the quality of service (QoS) factors such as throughput, delay, and jitter~\cite{firoiu2002qos}.
Eventually, QoS factors gave way to a more sophisticated measurement of quality of experience (QoE)~\cite{chen2014qos}, where users were directly asked about their experience while watching video content, typically via controlled laboratory experiments. 
Here, user sensitivities were an inextricable part of the evaluation but mixed with every other external factor that cannot be gauged by analyzing session data, i.e., confounding factors, such as interest in the particular content or testing conditions. 
Instead of separating these confounding factors, initial research focused on modeling the subjective QoE via objective metrics such as video bitrate or number of stalling events~\cite{robitza2018p1203}.
Finally, researchers realized the importance of tracking and modeling actual user engagement expressed in terms of view time~\cite{dobrian2011understanding} or user abandonment likelihood~\cite{lebreton2020predicting}. 
This measure has the critical advantage of directly impacting video streaming stakeholders such as service and/or content providers. 
However, collecting user engagement through controlled experiments is a significant challenge since viewers must simulate their watching patterns without choosing the provided content. 
Therefore, it is harder to single out user sensitivity from confounding factors. 
Although many studies have collected and utilized large datasets containing key features, none are in the public domain~\cite{dobrian2011understanding,balachandran2013developing,shafiq2014network}.
At the same time, gathering such a dataset is a tremendous effort, which requires users to install plugins on their systems that can record their interactions with streaming services. 
This is typically collected via crowdsourcing~\cite{robitza2020you} and following many users for weeks or months. 
Furthermore, it is complicated to track users across multiple devices. 
An alternative is collaborating with a video streaming provider to collect the necessary data~\cite{dobrian2011understanding}. 

\textbf{Digital Twins.}
Having established the importance of modeling user sensitivities and their impact on engagement, it is necessary to identify the best model for this task. 
A promising concept for modeling individual user sensitivities is that of digital twins (DTs), i.e., digital replicas of physical twins (PTs) that behave as closely as possible to the real entity for the intents and purposes of the application. 
This is an already established technology in the manufacturing sector, where it enhances monitoring and forecasting of machines’ conditions~\cite{tao2018digital}, but it struggles to find a broader scope of application, partly due to the difficulty of producing a general, application-agnostic definition of a DT. \cite{tao2018digital}. 
Therefore, the present study leverages DTs that focus on two properties: 
\begin{itemize}[leftmargin=*,noitemsep,topsep=0pt]
    \renewcommand{\labelitemi}{\scriptsize$\blacksquare$}
    \item Composition: a DT is not stand-alone but included in a system, together with its PT and an application that leverages the DT to understand or predict properties of the PT.
    \item Data persistency: a DT keeps a record of its PT states at all times.
\end{itemize}
In its most refined version, a video streaming engagement DT would replicate the exact behavior of its PT for that session, as is the case for other industries~\cite{moreno2017virtualisation}. 
In other words, it would pause, seek, stop watching, etc., when the user would. 
This DT is tough to reproduce in a video streaming setting, as fine-grained user behavior is often not related to the session itself but to confounding factors.

In this paper, we introduce DIGITWISE, a machine learning-based system that trains on data collected from every step of the video streaming pipeline, then leverages XGBoost-based DT models to learn unique user sensitivities, and outputs via another XGBoost model, predictions for video streaming engagement. 
Figure \ref{fig:system-overview} delineates such context. 
In addition to accurately predicting user engagement, DIGITWISE allows a system analyzer to explore how changing the value of a streaming parameter would impact a group of users’ engagement. 
All that is required is to generate several viewing sessions that differ only in a considered parameter and input these sessions into the DT model, which will output the engagement for each alternative. 
For example, a video streaming provider interested in the best encoding parameters for their videos can use this system to check the impact of each encoding parameter on their subscribers.

This paper makes the following contributions:
\begin{itemize}[leftmargin=*,noitemsep,topsep=0pt]
    \renewcommand{\labelitemi}{\scriptsize$\blacksquare$}
    \item We present new findings on the impact of viewers’ unique sensitivities on their video streaming engagement based on real-world streaming sessions.
    \item We release our extensive dataset of user streaming sessions both for the reproducibility of the present work and to allow further work in this respect, given the lack of publicly available data\footnote{Available at \hyperlink{https://github.com/emanuele-artioli/digitwise}{https://github.com/emanuele-artioli/digitwise}}.
    \item We develop DIGITWISE, a machine learning-based model that accurately estimates user engagement by leveraging the concept of DTs and user sensitivities.
    \item We utilize DIGITWISE in conjunction with a cloud-based testbed to compare the impact of different design choices (ABR, bitrate ladder, etc.) on user engagement.
\end{itemize}

\section{Related Work}

\begin{table*}[h]
\centering
\caption{Comparing the dataset of DIGITWISE with the state-of-the-art.}
\resizebox{\textwidth}{!}{%
\begin{tabular}{@{}lcccccccc@{}}
\toprule
               & Dobrian & Balachandran & Shafiq & Rodriguez & Robitza & Lebreton & Huang  & \textbf{DIGITWISE} \\
 &
  et al.\cite{dobrian2011understanding} &
  et al.~\cite{balachandran2013developing} &
  et al.~\cite{shafiq2014network} &
  et al.~\cite{rodriguez2014content} &
  et al.~\cite{robitza2020you} &
  et al.~\cite{lebreton2020predicting} &
  et al.~\cite{huang2022personalized} &  \\ \midrule
startup delay  & \checkmark  & \checkmark   &            & \checkmark & \checkmark  & \checkmark   & \checkmark & \checkmark             \\ \hline
bitrate        & \checkmark  & \checkmark   & \checkmark & \checkmark & \checkmark  & \checkmark   & \checkmark & \checkmark             \\ \hline
stalls         & \checkmark  & \checkmark   &            & \checkmark & \checkmark  & \checkmark   & \checkmark & \checkmark             \\ \hline
device type    &             & \checkmark   & \checkmark &            &             &              &            & \checkmark             \\ \hline
content type   & \checkmark  &              &            & \checkmark &             & \checkmark   & \checkmark &                        \\ \hline
video duration & \checkmark  &              & \checkmark &            & \checkmark  &              &            & \checkmark             \\ \hline
codec          &             &              & \checkmark & \checkmark &             & \checkmark   &            &                        \\ \hline
time of day    &             &              & \checkmark &            & \checkmark  &              &            & \checkmark             \\ \bottomrule

\end{tabular}%
}
\label{tab:related-dataset}
\end{table*}

\begin{table*}[h]
\centering
\caption{Comparing the architecture of DIGITWISE with the state-of-the-art.}
\resizebox{\textwidth}{!}{%
\begin{tabular}{lcccccccc}
\toprule
       & Dobrian           & Balachandran  & Shafiq        & Rodrıguez       & Robitza       & Lebreton        & Huang           & \textbf{DIGITWISE} \\
 &
  et al.\cite{dobrian2011understanding} &
  et al.~\cite{balachandran2013developing} &
  et al.~\cite{shafiq2014network} &
  et al.~\cite{rodriguez2014content} &
  et al.~\cite{robitza2020you} &
  et al.~\cite{lebreton2020predicting} &
  et al.~\cite{huang2022personalized} &
   \\ \midrule
input  & real sessions     & real sessions & real sessions & laboratory test & real sessions & laboratory test & laboratory test & real sessions      \\ \hline
model  & linear regression & decision tree & decision tree & ad-hoc function & N/A           & ad-hoc function & ad-hoc function & XGBoost            \\ \hline
output & engagement        & engagement    & engagement    & QoE             & engagement    & engagement      & QoE             & engagement         \\ \bottomrule
\end{tabular}
    }
    \label{tab:related-models}
\end{table*}

In this section, we investigate a set of studies that tackled the problem of modeling user experience in video streaming. 
The current state of the art can be categorized regarding the following features: 
($i$) the source of input data: real user sessions vs. controlled laboratory experiment,
($ii$) the employed model: ad-hoc, closed-form functions vs. machine learning models, and 
($iii$) the model's objective: QoE vs. engagement.
Dobrian et al.~\cite{dobrian2011understanding} provide foundational knowledge regarding the importance of moving toward modeling engagement and how different video streaming features impact it. 
They collect and analyze a unique dataset of client-side measurements from millions of video viewing sessions, focusing on different quality metrics and engagement levels (see Table~\ref{tab:related-dataset} for more details). 
Then, they apply correlation and linear regression of compressed metrics (e.g., number of buffering events, average quality) to model user engagement. 
Their findings have important implications for content providers to optimize user engagement by reducing buffering, minimizing buffering events for long video-on-demand and live content, and increasing the average bitrate for live content. 
The authors stress the significance of using measurement-driven insights to improve Internet video delivery. 
Balachandran et al.~\cite{balachandran2013developing} expand on the previous paper by using the same dataset to develop a decision tree able to classify user sessions into 10\% bins of engagement with 70\% accuracy. 
To achieve this, they analyze potential confounding factors, removing or dividing them into different classes before training, allowing the model to avoid many pitfalls. 
Shafiq et al.~\cite{shafiq2014network} investigate the impact of network factors on user engagement. 
They build a bagged decision tree-based model capable of accurately predicting whether a user will stop watching before reaching the end of the video by only utilizing the first 10 seconds of network activity. 
Rodriguez et al.~\cite{rodriguez2014content} investigate the importance of the user's preference for video content and its impact on a {\color{black}subjective} video quality metric. 
They classify content into soccer, reportage, and tech documentary and assign each user a coefficient per content. 
These coefficients are added as variables to typical streaming features like rebuffering in a preference factor function that resizes the opinion score of each view. 
Another valuable contribution of this paper is the usage of video lengths similar to real streams.
Robitza et al.~\cite{robitza2020you} use crowdsourcing to collect an extensive dataset about user engagement on multiple popular services, with the vast majority of recorded sessions coming from YouTube. 
By leveraging the ITU-T P.1203 QoE model~\cite{Raake2017} and session-level information, they show commonalities and differences among user groups and how these relate to whether users stopped watching at the beginning, middle, or end of a streaming session.
Lebreton et al.~\cite{lebreton2020predicting} consider a more extensive set of features and use them to model the user quitting ratio as a function of time. 
They offer streaming platforms the ability to predict the likelihood of losing a certain number of viewers due to encoding and ABR decisions. 
They ponder using user sensitivities as a feature to aid in modeling engagement but ultimately limit their scope to model users quitting because of bad quality and not user engagement.
Huang et al.~\cite{huang2022personalized} realize the potential of leveraging DT-based solutions for video streaming user experience. 
To develop an optimal ABR algorithm, they extract user QoE factor sensitivities through nonlinear regression. 
Having obtained the set of factor coefficients for each user, they compute a PQoE model that considers the impact of factors exponentially decreasing with time.
Tables~\ref{tab:related-dataset} and \ref{tab:related-models} compare the proposed DIGITWISE and related work on two aspects: the completeness of the dataset and the model architecture. 
The dataset used is the most complete with respect to features that have been shown to impact user engagement. 
Model-wise, DIGITWISE does not include the content type and codec simply because those features were missing from the training data due to our data provider's privacy policies. 
A more complete dataset would not require changes in the DIGITWISE architecture and would allow a more robust modeling of user engagement.

\section{DIGITWISE}

\begin{figure}
    \centering
    \includegraphics[width=\linewidth]{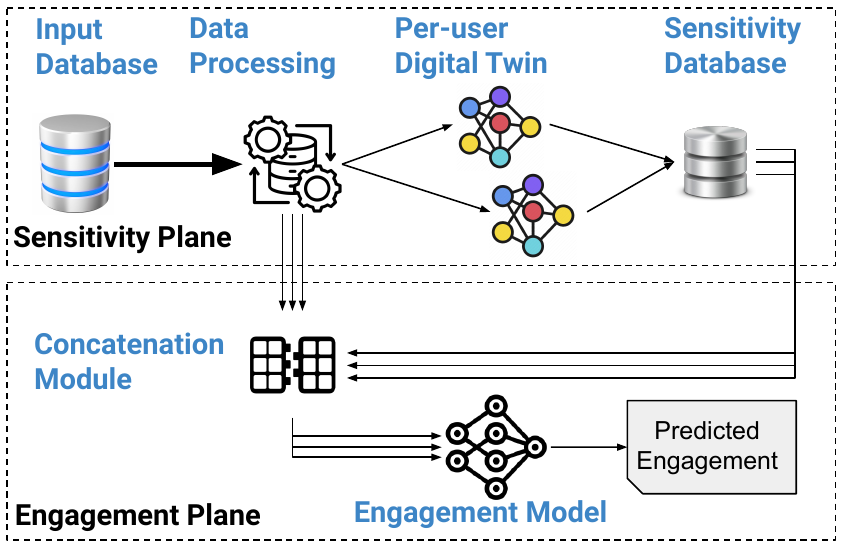}
    \caption{DIGITWISE architecture.}
    \label{fig:model-architecture}
\end{figure}

\begin{figure}
    \centering
    \includegraphics[width=.8\linewidth]{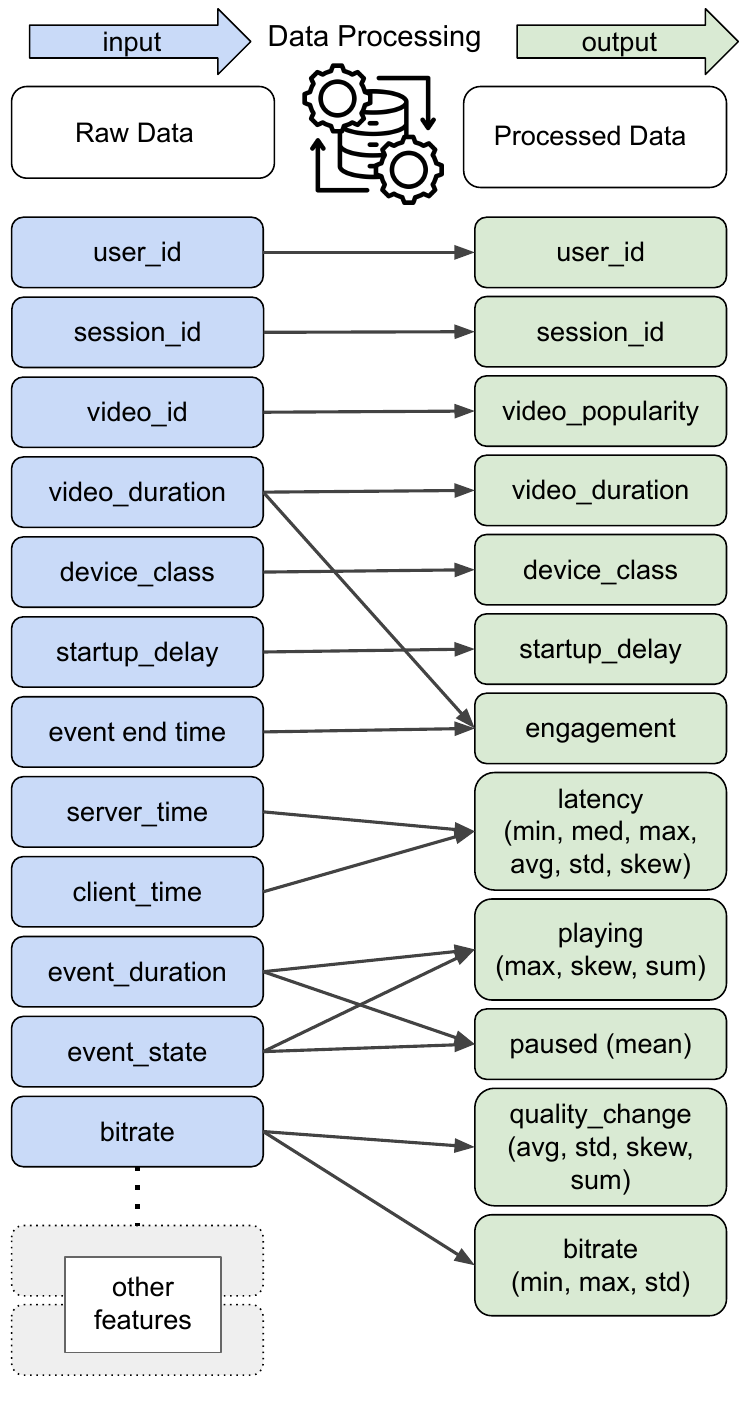}
    \caption{Inputs and outputs of the Data Processing component.}
    \label{fig:input_output}
\end{figure}


Figure 2 illustrates the DIGITWISE architecture, including the two main planes: ($i$) \textit{sensitivity plane} and ($ii$) \textit{engagement plane}.

The \textbf{sensitivity plane} consists of four components: input database, data processing, per-user DT, and sensitivity database. 
The process begins by collecting data from Bitmovin Analytics~\cite{analytics} into a database. 
In this study, the dataset consists of one row per session event, meaning a new row of data is collected each time a user starts, plays, pauses, seeks, or quits the session or an internal event happens during the streaming that impacts QoE (e.g., bitrate change, stall). 
This new row of data includes information about time, location, user identifier (ID), video ID, session ID, device type, bitrate, etc. 
Such a rich dataset allows us to draw accurate conclusions about the impact of many factors on user engagement, isolate the effects of confounding factors, and verify the results of changing the values of a factor.
Subsequently, the data undergo a processing task aimed at:
($i$) eliminating sessions, i.e., the rows relating to the same viewing experience, with collection errors or missing data, 
($ii$) engineering new useful features such as video popularity or engagement, 
($iii$) consolidating all data related to the same session into a single row through the calculation of aggregate metrics, 
($iv$) selecting users with enough data to enable reliable models, and 
($v$) performing feature selection. 
Figure~\ref{fig:input_output} shows the input and output parameters to and from the data processing component.
The processed data is then utilized to train a set of initial models known as DTs. 
Each DT is trained on a subset of sessions by a specific user, enabling the models to learn individual user sensitivities. 
Once the models are effectively trained, information gain~\cite{shannon1948entropy} and permutation importance~\cite{altmann2010permutation}, which will be explained in the following subsection, are employed to extract user sensitivities. 
These sensitivities, which provide valuable insights into user engagement, replace the need for user IDs since they contain information regarding what differentiates one user from others, i.e., its sensitivities becoming available in the \textit{sensitivity database}.\par 
The \textbf{engagement plane}, consisting of two main components (i.e., concatenation module and engagement model), uses the provided data by the sensitivity plane to estimate user engagement accurately. 
The \textit{concatenation module} combines the provided sensitivity features with the selected features associated with the same user from the data processing component. The output of the concatenation module is fed into a unified \textit{engagement model}. 
The engagement model is trained on all user-engineered features and their sensitivities. 
Consequently, it can leverage similarities among all users while providing distinct engagement predictions based on the individual user.
In the following, we dive into the details of the sensitivity and engagement planes.

\subsection{Sensitivity Plane}

\textbf{Input Database.}
Data about user sessions, collected from player logs by the analytics server, are stored in a database.
Each record includes more than 100 raw features that capture all events occurring in the player, whether caused by the user or internal player processes (e.g., ABR algorithm). 
These features encompass information such as the time of day the video was streamed, what content delivery network (CDN) server was used to deliver it, what user and device streamed it, and when events like bitrate switch and stalls occurred.

\begin{figure*}[h]
    \centering
    \includegraphics[width=\textwidth]{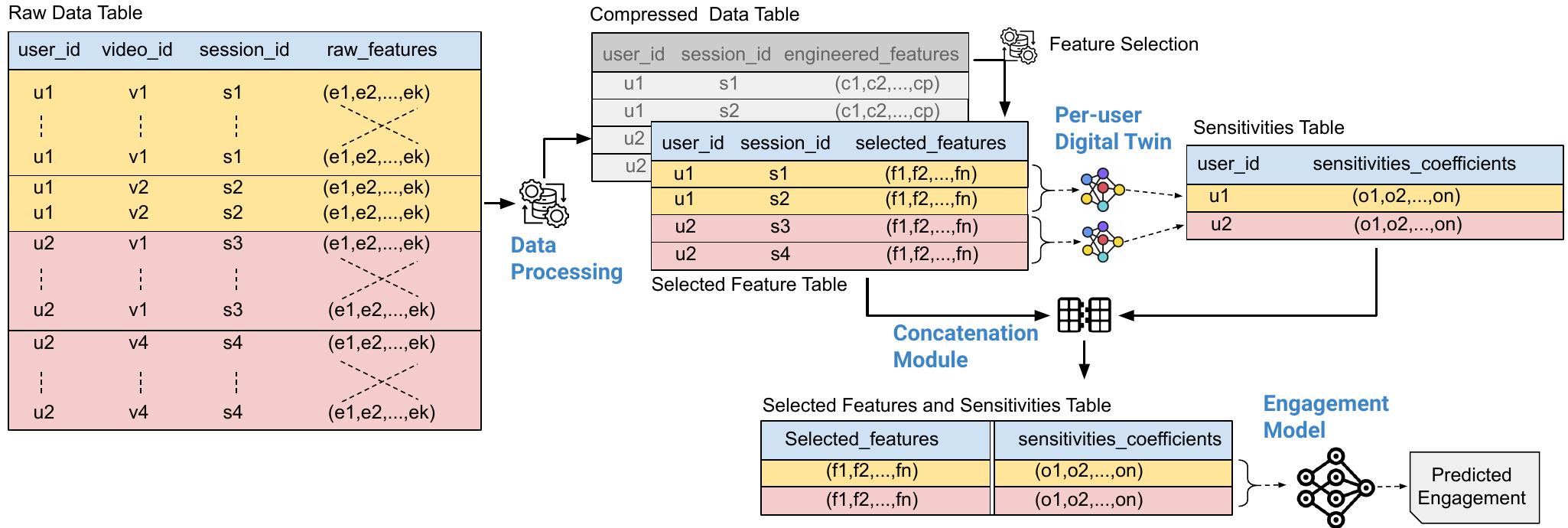}
    \caption{\textcolor{black}{DIGITWISE pipeline.}}
    \label{fig:DIGITWISE_pipeline}
\end{figure*}

\textbf{Data Processing.} 
Figure~\ref{fig:DIGITWISE_pipeline} shows the DIGITWISE pipeline consisting of various tables given to/generated by computational components. 
The \textit{Raw Data} table shows the prototype of raw data including $3+k$ features: \texttt{user\_id}, \texttt{video\_id}, \texttt{session\_id}, and $k$ extra raw features.
The first step in our preprocessing pipeline regards data cleaning. 
Many collection errors and incongruent information plague the raw features, therefore we set up a series of commands to identify and fix these issues: 
\begin{itemize}[leftmargin=*,noitemsep,topsep=0pt] 
\renewcommand{\labelitemi}{\tiny$\blacksquare$}
    \item Standardize null values into a value recognized by the models;
    \item Remove users with watching patterns incompatible with real, single people, i.e., watching the same video hundreds of times;
    \item Remove users with less than 50 viewing sessions, to avoid training DTs on too little data;
    \item Remove users who watched less than five different videos, to avoid picking up on patterns that are specific to the video and not the user;
    \item Video duration is not consistent along each video id, so it is standardized to each video's max video duration;
    \item Remove sessions that encountered errors, since we only need sessions that ended because of the user's choice;
    \item Remove sessions with negative event end time, which would make it impossible to gauge engagement;
    \item Remove sessions that were watched for less than 10 seconds;
    \item Remove sessions longer than 24 hours since they are likely to be 24/7 live channels whose engagement we cannot compare to regular videos;
    \item Event duration is sometimes negative, thus, remove affected sessions;
    \item Sometimes videotime end is greater than video duration, which is impossible, thus, remove affected sessions.
\end{itemize}
We also take care of engineering several features to allow the models to extract as much information about the viewing sessions as possible:
\begin{itemize}[leftmargin=*,noitemsep,topsep=0pt]
\renewcommand{\labelitemi}{\tiny$\blacksquare$}
    \item Convert time into hour of day, as time of day impacts user engagement~\cite{6883588};
    \item Convert video id into a video popularity score, that reflects how many times each video was watched, which is another feature impacting user engagement~\cite{6883588};
    \item The type of event (play, pause, seek, etc.) gets condensed into a more precise and less redundant space;
    \item Device class (TV, laptop, phone, etc.) gets remapped into screen size, to leverage its order;
    \item Add bitrate switches as the difference between the bitrate in the previous event and the current event;
    \item Add latency between client and analytics server as the difference between the timestamp added at client side before sending data and the one collected at server side as data arrives;
    \item Engagement is derived as maximum videotime end divided by video duration, in other words the furthest point in the video the user reached before quitting.
\end{itemize}

\textbf{Data Compressing.}
The analytics server collects data from one row per event, meaning that each time the player receives a playback request, starts playing, changes bitrate, pauses, etc., one new row is appended to the data about that user, video, and session. 
This data shape is valuable for time-sensitive applications such as an online DT predicting real-time engagement. 
Our current goal of offering video streaming service providers insight into their design choices will likely not benefit from the added complexity of developing a time-sensitive model to handle event-based data.
Therefore, the raw data needed to be compressed from one row per event into one row per user/session via aggregate metrics such as averages and standard deviations. 
As shown in Figure~\ref{fig:DIGITWISE_pipeline}, the occurred events during two sessions of user $u1$ are compressed into two rows in \textit{Compressed Data} table, each of which has $2+p$  features including  \texttt{user\_id}, \texttt{session\_id}, and $p$ \texttt{engineered\_features}, where $k<p$.
Compressing multiple rows of raw data into one row inevitably reduces some parts of information due to the order and positions of events that significantly impact user engagement. 
For example, this is clear when considering the scenario of two users playing the same video simultaneously. 
The first user experienced the first minute of playback with frequent stalls and is now at minute 10 after 9 minutes of smooth playback. 
The second user experienced the first 9 minutes without problems and is now overwhelmed by frequent stalls for a minute. 
Which user do we assume is the most likely to stop engaging with the content in the immediate future?
This problem is highlighted in~\cite{lebreton2020predicting}, and we offer a simple but efficient solution as follows.
To retain information on whether certain events were concentrated more towards the playback's beginning or end, we compute the skewness metric for features that benefit from it, such as bitrate switches and stalls. 
In the example above, while the stalls' mean and standard deviation are the same, the skewness is positive for the first user with initial stalls and negative for the second where the stalls occur at the end.

\textbf{User Selection.}
To avoid models that are skewed towards a particularly prevalent engagement value, we separate sessions into ten engagement bins (meaning, we group sessions with engagement from 0 to 10\%, from 10\% to 20\%, etc.) and sample these groups so that they have the same number of sessions. 
Then, we count how many sessions are left for each user and remove each user with less than 100 sessions to avoid DTs from training on too few data points to be relied upon. 
Once filtered, each user's sessions are split, 80\% make up a training set, the remaining a test set.

\textbf{Feature Selection}.
Another important step of the processing is feature selection. Many of the $p$ engineered features show a high correlation. 
Although not directly harmful to model predictions, a high number of features slows down training and hinders the production of feature importance scores. 
Therefore, we devise a semi-automatic feature selection comprising the following steps:
\begin{itemize}[leftmargin=*,noitemsep,topsep=0pt]
\renewcommand{\labelitemi}{\tiny$\blacksquare$}
    \item We train a Random Forest Regressor~\cite{ho1995random}
    model on the compressed dataset.
    \item From the trained model, we extract feature importance computed as the mean decrease in entropy~\cite{shannon1948entropy}.
    \item We calculate the Spearman rank correlation~\cite{myers2013correlation} coefficient matrix for the feature set.
    \item We sum the correlation coefficients for each feature to compute a rough indication of how much each feature is correlated to all others.
    \item We divide the previously obtained feature importance by the sum of correlation coefficients just computed to obtain a new measure of feature importance that is penalized when other features are strongly correlated to it. 
    \item For better interpretation, we normalize the coefficients.
\end{itemize}
The features with the highest coefficient are important and have low correlations among themselves. 
We can select the feature set that maximizes the performance to computing cost tradeoff by optimizing a single threshold for feature importance.
Furthermore, we removed any feature that might compromise the customer's or its users' privacy. 
The \texttt{selected feature table} in Figure~\ref{fig:DIGITWISE_pipeline} shows $2+n$ columns including \texttt{user\_id}, \texttt{session\_id}, and $n$ selected features, where $n<p$.
It is important to note that determining an appropriate threshold value is a delicate and fundamental tradeoff: a high threshold value will only allow the few most significant features, resulting in a fast, simple, and interpretable model.
However, the model runs the risk of missing out on the performance gain that less important features might have brought. 
A low threshold allows the model to extract as much information as possible from the data but increases its computational requirements and might even hinder its performance in the case of collinear features, as explained in the previous section.

\textbf{Per-user DT.} For each user's processed session data, e.g., two rows associated with user $u1$ in the \textit{selected feature table} in Figure~\ref{fig:DIGITWISE_pipeline}, we train a model, taking the role of a DT and modeling the user features into an engagement prediction. 
In other words, we need to determine the sensitivity coefficient value for each user's selected feature based on the calculated engagement value from the raw data (see the \texttt{sensitivities table} in Figure~\ref{fig:DIGITWISE_pipeline}). 
For this task, we use XGBoost~\cite{chen2016xgboost}, an ensemble algorithm that leverages boosted decision trees to achieve high performance by focusing its training on the iteratively hardest predictions. 
Aside from its ease of interpretation and fast convergence, literature has shown how XGBoost is often among the best predictors of tabular data~\cite{fernandez2019extensive}.
It is worth mentioning that the optimal hyper-parameters for each user's model, such as the number of trees to train and the number of features each tree had at its disposal, are identified through halving grid search~\cite{bergstra2012random}. 
This technique generates many candidate models with randomly picked sets of hyper-parameters, trains them on limited amounts of data, evaluates their performance via cross-validation, discards the ones with poor performance, and trains the survivors on more data, iterating until only the best model is left.
Having obtained a reliable level of performance, these models are scrutinized to extract each feature's impact on engagement. 
XGBoost natively offers the importance of features via information gain~\cite{shannon1948entropy}.

\subsection{Engagement Plane}
\textbf{Concatenation Task.} Having obtained each feature's importance for each user, we use this information about user sensitivities as an embedding of that user, substituting the sensitivities to the user ID (see \texttt{selected feature and sensitivities table} in Figure~\ref{fig:DIGITWISE_pipeline}). 
Category embedding~\cite{csenel2018semantic} is a typical technique in machine learning that aims at compacting a high-cardinality categorical feature, i.e., user ID, into a low-dimensional, numerical vector where each scalar represents that user's position concerning other users in a particular dimension.
This step allows machine learning models to extract meaning from categorical features, which would be hard without converting them into embedded numerical dimensions.
These dimensions are usually hard to interpret and are discovered by the embedding algorithm upon training. 
On the contrary, each dimension of our user sensitivity vector has a precise meaning, i.e., it represents the sensitivity of that user to a specific selected feature.

\textbf{Engagement Model.} These sensitivities, which provide valuable insights into user engagement, replace the need for user IDs since they contain all the information regarding what differentiates one user from others. 
Once the user sensitivities have been concatenated with their respective sessions, these enhanced sessions are fed as input into a unified XGBoost model.
This model is optimized via halving grid search~\cite{bergstra2012random}, similar to the sensitivity models. 
This way, we get the best of both worlds: a sufficient amount of sessions to extract the general behavior (similar to a degree in each user) and the digital-twin improvement upon it by using specific information about each user.

\section{Performance Evaluation}
To evaluate the performance of DIGITWISE in various scenarios and also compare it with state-of-the-art approaches, we set up a Python Jupyter notebook to run on an Ubuntu server sporting an Intel Xeon Gold 5218 CPU @ 2.30GHz (128 cores), an NVIDIA Quadro GV100 GPU (32 GB VRAM, 5,120 CUDA Cores), 384 GB of RAM, and 9.6 TiB of RAID5 SSD storage. 
The code is available on GitHub, together with the data used for the training\footnote{Available at \hyperlink{https://github.com/emanuele-artioli/digitwise}{https://github.com/emanuele-artioli/digitwise}}.

\subsection{Dataset}
In this study, we focus on one dataset consisting of a real world video streaming services related to car race events between May and August 2023. 
This translated into a dataset with 102,233,138 rows and 102 features, describing events from 3,261,381 unique viewing sessions of 50,942 real users watching 3,181 different videos. 
Such a varied dataset contains much information deemed irrelevant to the current study. 
Therefore, only the features shown in Figure~\ref{fig:input_output} are effectively used by DIGITWISE. 
Vice versa, many features that are useful for engagement modeling are absent from the available data. 
Notable examples include the codec, any information about the session's environment such as viewing distance or light conditions, user demographics, etc. 
Furthermore, the data relates to a single content provider with a narrow scope (car races). 
Therefore, considerations about content type and impact on the user base are impossible.
Including greater variability and additional features will likely improve DIGITWISE's performance, and future efforts will be made in this direction.

\subsection{Experimental Scenarios}

We define the following scenarios to assess DIGITWISE's performance under varying design parameters. 
Additionally, we will compare it with three alternative approaches. 
Finally, we will demonstrate how DIGITWISE can assist video streaming service providers to fine-tune their streaming service parameters.

\textbf{Scenario 1}: \textit{Comparing the impact of  XGBoost and Multi-Layer Perceptron (MLP) on the DIGITWISE performance.}
We identified two potential model architectures for DIGITWISE, namely XGBoost and MLP~\cite{muller1995neural}. 
These represent the two leading approaches with regard to machine learning on tabular data, respectively, Decision Tree-based models and Neural Networks. 
Both models are optimized via halving grid search, and their training time and mean average error results will be compared.
We are interested in a model that is not only precise in its prediction of engagement, but can also offer accurate predictions early in the session. 
Customers would benefit from knowing the user engagement after only a few minutes into the session, and be thus able to implement changes in the session before the user stops engaging. 
Therefore, each model is trained multiple times, each time with a different training dataset where only session events happening in the first seconds or minutes are included. 

\textbf{Scenario 2}: \textit{The impact of extracting sensitivities coefficients on the DIGITWISE performance.}
After identifying the optimal data modeling architectures, we assess the influence of user sensitivities. 
This evaluation involves training the selected model architectures on the same data twice: first without considering user sensitivities and then with the inclusion of user sensitivities. 
The discrepancy in performance between these two models reflects the additional information gained from incorporating the sensitivity coefficients.

\textbf{Scenario 3}: \textit{Determining the appropriate threshold value for the feature selection module.}
We mentioned before how a single threshold value is responsible for the feature selection step. 
By tweaking this value, we can influence which features will make their way into the training data for DIGITWISE. 

\textbf{Scenario 4}: \textit{Assessing the performance of DIGITWISE compared to the state-of-the-art approaches.}
In this scenario, we will compare the performance of DIGITWISE with the following state-of-the-art models over various time horizons and threshold values: 
\begin{itemize}[leftmargin=*,noitemsep,topsep=0pt]
\renewcommand{\labelitemi}{\scriptsize$\blacksquare$}
    \item Balachandran et al.~\cite{balachandran2013developing} sets up its Decision Tree-based model for a classification task of predicting user engagement in a 10-bin fashion using the whole session data. 
    They obtain an accuracy of 70\%. In other words, 7 out of 10 times, the model classifies the real session engagement within a margin of error of 10\%.
    \item Shafiq et al.~\cite{shafiq2014network} developed a binary model capable, with 87\% accuracy, of predicting whether the user will download the whole video after only seeing the first 10 seconds of network data.
    \item Finally, we compared DIGITWISE with~\cite{lebreton2020predicting}, whose model groups the data by video and outputs the percentage of users that will stop watching each video "midway through". They then compute the Pearson Correlation Coefficient~\cite{myers2013correlation}. 
\end{itemize}

\textbf{Scenario 5}: \textit{Changing streaming parameters and measuring its impacts on the user engagement by DIGITWISE.}
In this scenario, we will show a useful application of the DIGITWISE model in fine-tuning the streaming parameters by measuring user engagement. 
To this end, we leverage CAdViSE~\cite{taraghi2020cadvise}, a cloud-based adaptive video streaming evaluation framework for the automated testing of adaptive media players. 
CAdViSE simulates content streaming using Amazon Web Services (AWS). 
Its main goal is to offer a testbed for comparing different streaming parameters (e.g., ABR algorithms, bitrate ladder, etc.) under the same network and player conditions, but it is used in this context to produce realistic session data whose parameters can be carefully determined.

\begin{figure}
    \centering
    \includegraphics[width=\linewidth]{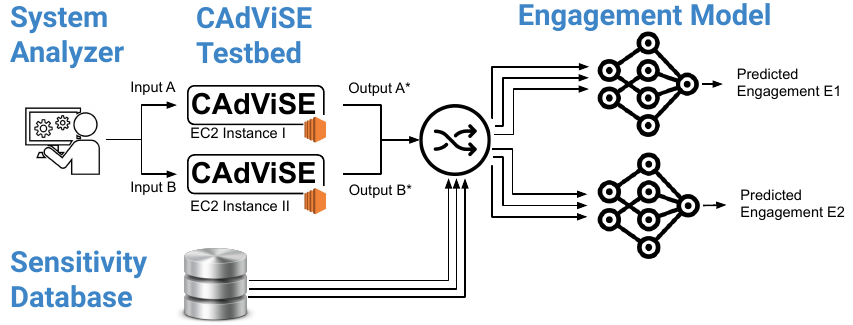}
    \caption{Service tuning architecture.}
    \label{fig:model-service-tuning-architecture}
\end{figure}

Figure~\ref{fig:model-service-tuning-architecture} shows the architecture used for Scenario~5. For example, let's assume a system analyzer is interested in the impact that a particular ABR algorithm has on a set of users.
They can then generate CAdViSE sessions that share the same network trace, content streamed, segment size, etc., and differ only in their ABR of choice. 
By adding the interested users' sensitivities to these simulated sessions and using the combination as input to the trained engagement model, the system analyzer can obtain multiple engagement predictions for each user, one prediction per ABR, and their difference is the expression of the impact of said ABR on that user's engagement.

To benchmark this scenario, we generate 100 CAdViSE sessions for each combination shown in Table~\ref{tab:cadvise-distribution}.
We concatenate the generated sessions with user sensitivities drawn randomly from the available users and run the engagement model to predict an engagement value for each session. Then, we compute aggregate metrics such as the average engagement for each value of the initially chosen feature.

\begin{table}[h]
\centering
\caption{Measuring the engagement over various CAdViSE sessions distribution.}
\resizebox{\columnwidth}{!}{
\begin{tabular}{@{}lcllllll@{}}
\toprule
Segment size (s) & ABR                           & Network trace              & Mean  & Std.  & Min   & Median & Max   \\ \midrule
1                & L2A                           & AmazonFCC                  & 0.638 & 0.123 & 0.471 & 0.720  & 0.788 \\ \midrule
1                & L2A                           & Cascade-20                 & 0.616 & 0.127 & 0.455 & 0.576  & 0.785 \\ \midrule
1                & L2A                           & Cascade-5                  & 0.611 & 0.142 & 0.415 & 0.575  & 0.786 \\ \midrule
1                & L2A                           & Constant (16 Mbps)         & 0.631 & 0.124 & 0.403 & 0.685  & 0.763 \\ \midrule
1                & L2A                           & LTE~\cite{vanderHooft2016} & 0.634 & 0.132 & 0.425 & 0.576  & 0.785 \\ \midrule
2                & Bola                          & AmazonFCC                  & 0.597 & 0.145 & 0.411 & 0.567  & 0.785 \\ \midrule
2                & Bola                          & Cascade-20                 & 0.674 & 0.120 & 0.453 & 0.744  & 0.785 \\ \midrule
2                & L2A                           & AmazonFCC                  & 0.601 & 0.140 & 0.423 & 0.569  & 0.783 \\ \midrule
2                & L2A                           & Cascade-20                 & 0.588 & 0.138 & 0.412 & 0.567  & 0.785 \\ \midrule
2                & L2A                           & Cascade-5                  & 0.627 & 0.136 & 0.428 & 0.576  & 0.786 \\ \midrule
2                & L2A                           & Constant (16 Mbps)         & 0.638 & 0.137 & 0.451 & 0.744  & 0.771 \\ \midrule
2                & L2A                           & Constant (4 Mbps)          & 0.610 & 0.123 & 0.468 & 0.554  & 0.757 \\ \midrule
2                & L2A                           & LTE                        & 0.554 & 0.127 & 0.398 & 0.565  & 0.687 \\ \midrule
2                & LoL$+$~\cite{lim2020they}     & Cascade-5                  & 0.621 & 0.145 & 0.427 & 0.572  & 0.786 \\ \midrule
2                & LoL$+$                        & LTE                        & 0.620 & 0.137 & 0.423 & 0.569  & 0.787 \\ \midrule
2                & Throughput~\cite{liu2011rate} & AmazonFCC                  & 0.610 & 0.139 & 0.400 & 0.573  & 0.784 \\ \midrule
2                & Throughput                    & Cascade-20                 & 0.634 & 0.127 & 0.453 & 0.722  & 0.770 \\ \bottomrule
\end{tabular}
}
\label{tab:cadvise-distribution}
\end{table}

This tells us the potential for that design choice: if we were hypothetically using the value for each feature that is shown to have the lowest engagement for the chosen users, how much could the engagement improve by switching to the value shown to have the highest engagement?

\subsection{Results and Analysis}

\textbf{Scenario 1}. 

\begin{figure*}[h]
\minipage{0.33\textwidth}
  \includegraphics[width=\linewidth]{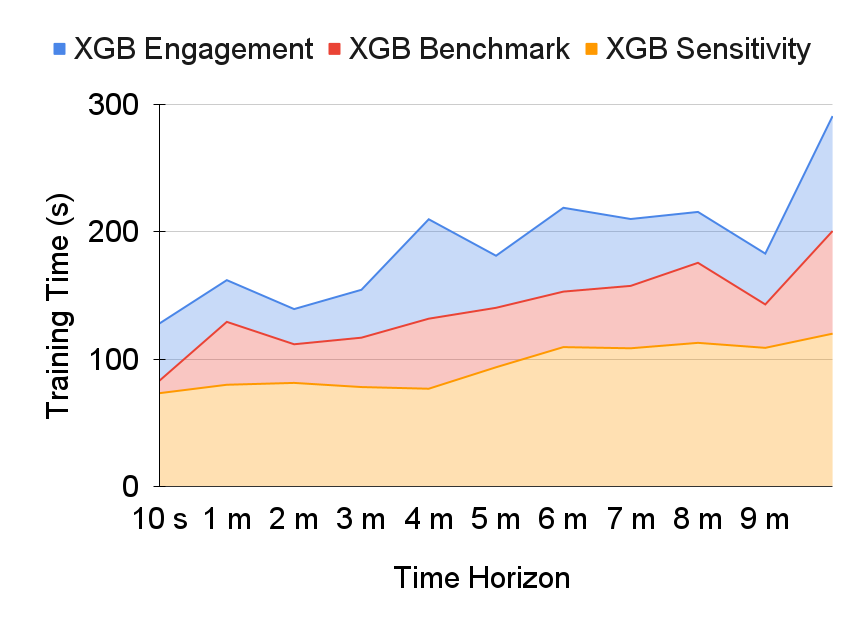}
  \caption{XGBoost training time.}\label{fig:xgb-training-time}
\endminipage\hfill
\minipage{0.33\textwidth}
  \includegraphics[width=\linewidth]{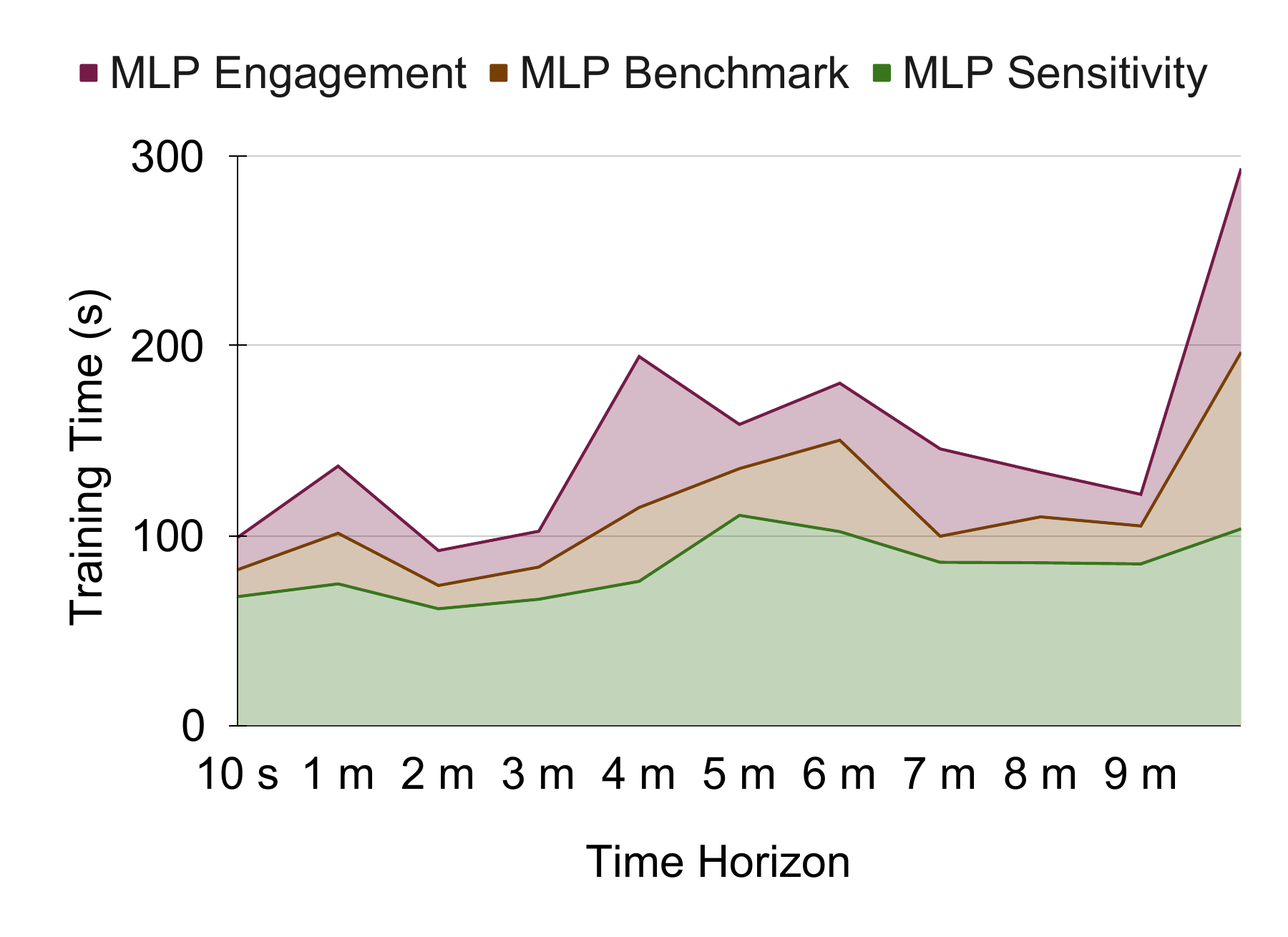}
  \caption{MLP training time.}\label{fig:mlp-training-time}
\endminipage\hfill
\minipage{0.33\textwidth}%
  \includegraphics[width=\linewidth]{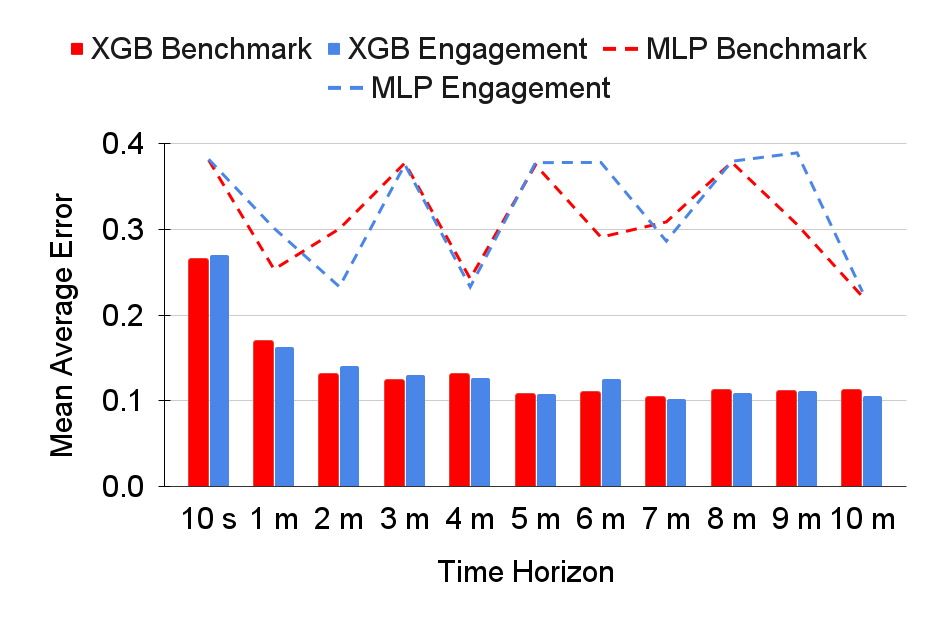}
  \caption{MAE comparison.}\label{fig:mae-comparison}
\endminipage
\end{figure*}

Figure~\ref{fig:xgb-training-time} and Figure~\ref{fig:mlp-training-time} show the cumulative times for the training of:
($i$) the sensitivity models, 
($ii$) the benchmark models, i.e., the models that train on the whole data set without the sensitivity features, and 
($iii$) the engagement models for XGBoost and Multi-Layer Perceptron.
The architectures require similar times to train, which is important to allow for a fair performance comparison. Furthermore, the figures also show how the time required to train depends not on the time horizon but on the model under training: for both architectures, most time is spent training the sensitivity model. 
Considering that one sensitivity model is required for each user while the same engagement model is used for all users, one can easily understand such behavior.

Regarding the Mean Average Error (MAE), as shown in Figure~\ref{fig:mae-comparison}, XGBoost obtains a much lower MAE when compared to MLP, regardless of available data. 
Furthermore, it shows a more consistent behavior, steadily improving its performance as more data is available, then settling around the 7-minute time-horizon mark, likely for a combination of two factors: diminishing returns of a longer time horizon and difficulty in lowering an already low MAE. 
This is not true for MLP, especially in its engagement phase. 
Further evaluation might shed light on this peculiar behavior of MLP, but is deemed irrelevant for the current study, which accepts XGBoost as the superior model for further performance evaluation and practical applications. 

\textbf{Scenario 2}. 

\begin{figure}[h]
    \centering
    \includegraphics[width=\linewidth]{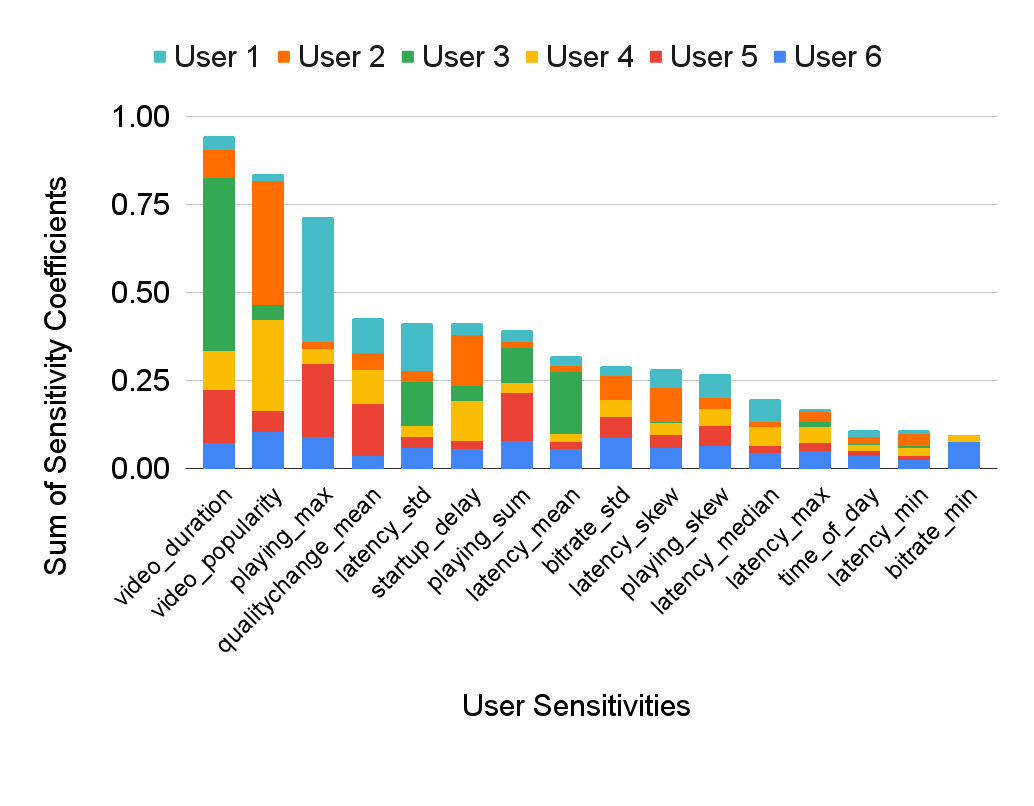}
    \caption{Model sensitivities.}
    \label{fig:model-sensitivities}
\end{figure}

Figure \ref{fig:mae-comparison} reports another important metric. 
We benchmark our model, i.e., XGB Engagement, against a model with the same architecture, trained on the same streaming session, and only lacking the information about user sensitivities, i.e., XGB Benchmark.
We show how personal user sensitivities can reduce the mean average error by up to $5.8$\% (for the model with a 10-minute time horizon) and, therefore, prove to be a determining factor in users' willingness to continue watching. 
Moreover, Figure~\ref{fig:model-sensitivities} shows the sensitivity of users to each training feature. 
For clarity, only users with a high number of sessions were selected.
User sensitivities show a clear trend, where most users typically share the most important features. Still, a significant degree of individuality can be observed.
Video duration and popularity are the most important factors in determining the final user engagement. 
This is not surprising, especially in the case of video duration, as a very long video requires much more interest from a user to be completed than a short one 
Furthermore, long videos have a much higher chance of being affected by network, external situations, and several different confounding factors that can cause a premature interruption in the session. 
It is important to note how this result is directly caused by how the engagement metric was measured: if engagement were measured as time elapsed before quitting, instead of a percentage of video duration, the first few minutes of a video would have the same chance for interruption regardless of duration. 
However, this would have rendered our study not comparable with the state-of-the-art and would have introduced other issues.
The most critical issue would have been the inversion of the aforementioned trend since a short video would be penalized by not having the chance to reach the watch time of a long one. 
Therefore, while not perfect, the chosen engagement metric was nonetheless deemed correct.
Another important observation to draw from Figure~\ref{fig:model-sensitivities} is how, while being by far the most important feature overall, no single feature is the most important for most users.
This is crucial as it represents the main reason for the present study. It demonstrates how users noticeably differ in their importance of video streaming features and how different factors influence their behavior. \par

\textbf{Scenario 3}. 

\begin{figure*}[h]
\minipage{0.33\textwidth}
  \includegraphics[width=\linewidth]{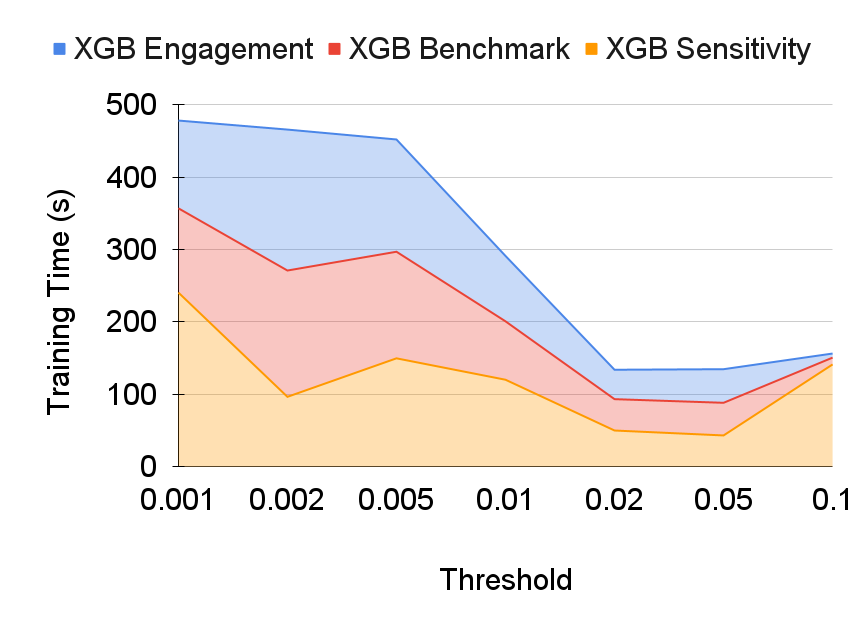}
  \caption{XGBoost training time over\\ different threshold values.}\label{fig:xgb-training-time2}
\endminipage\hfill
\minipage{0.33\textwidth}
  \includegraphics[width=\linewidth]{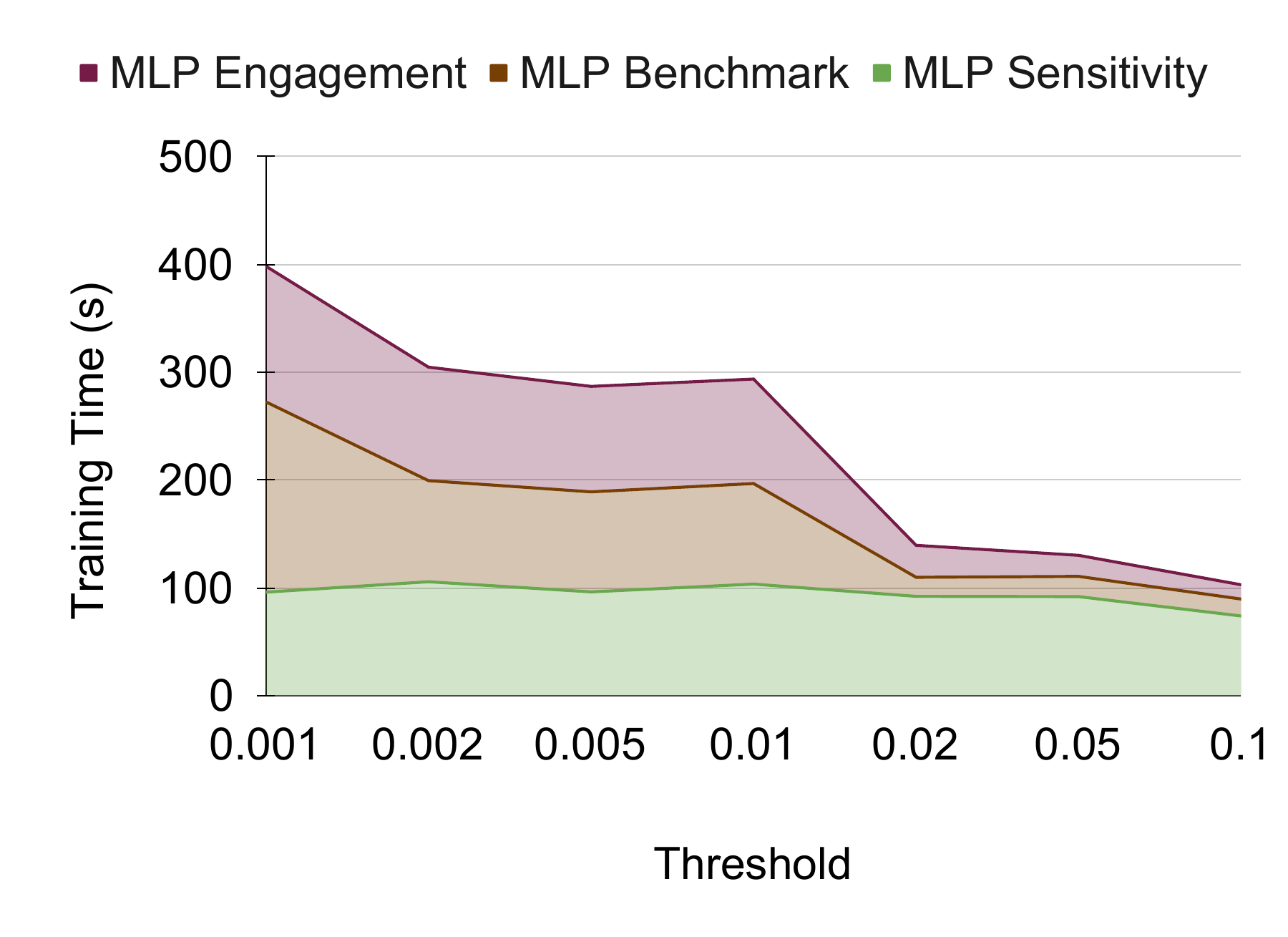}
  \caption{MLP training time over\\ different threshold values.}\label{fig:mlp-training-time2}
\endminipage\hfill
\minipage{0.33\textwidth}%
  \includegraphics[width=\linewidth]{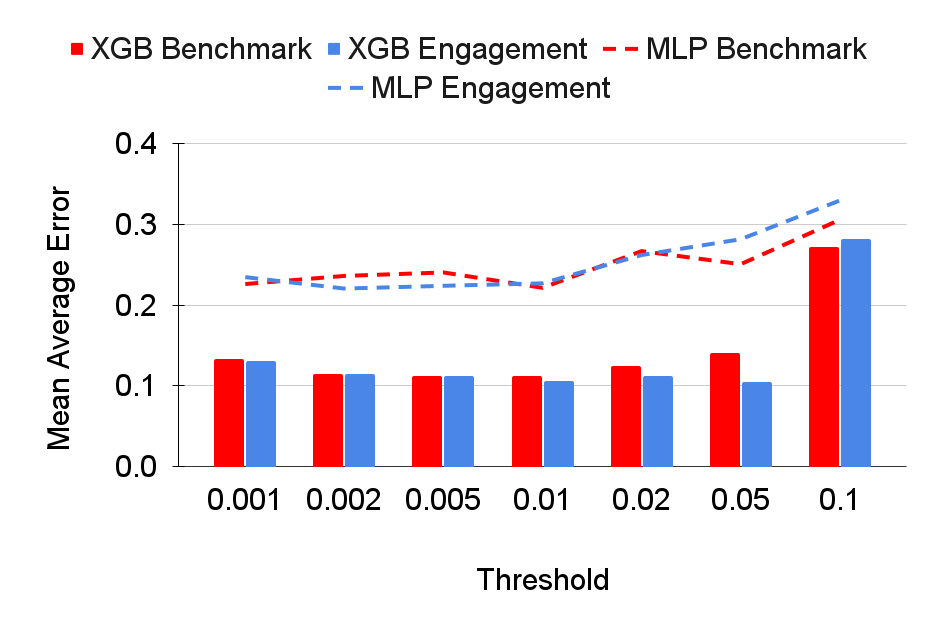}
  \caption{MAE comparison over different\\ threshold values.}\label{fig:mae-comparison2}
\endminipage
\end{figure*}

To identify the optimal feature selection threshold, we trained the model at various threshold levels, from very low, where almost every available feature was included, to very high, where only features of the utmost importance were selected. 
The results are shown in Figures~\ref{fig:xgb-training-time2}-~\ref{fig:mae-comparison2} .
As the figures show, very high thresholds do not allow enough features to offer correct predictions, while very low thresholds do not incur the aforementioned risk of collinearity. 
This is because the XGBoost model's architecture, like every decision tree-based architecture, is resistant to feature collinearity and can disregard unimportant features and focus on the important variability. 
However, Figure~\ref{fig:xgb-training-time2} demonstrates how the training time increases with lower thresholds because of the higher number of features, and the model interpretation is likewise harder. 
\par

\textbf{Scenario 4}. 

\begin{figure*}[h]
\minipage{0.5\textwidth}
  \includegraphics[width=\linewidth]{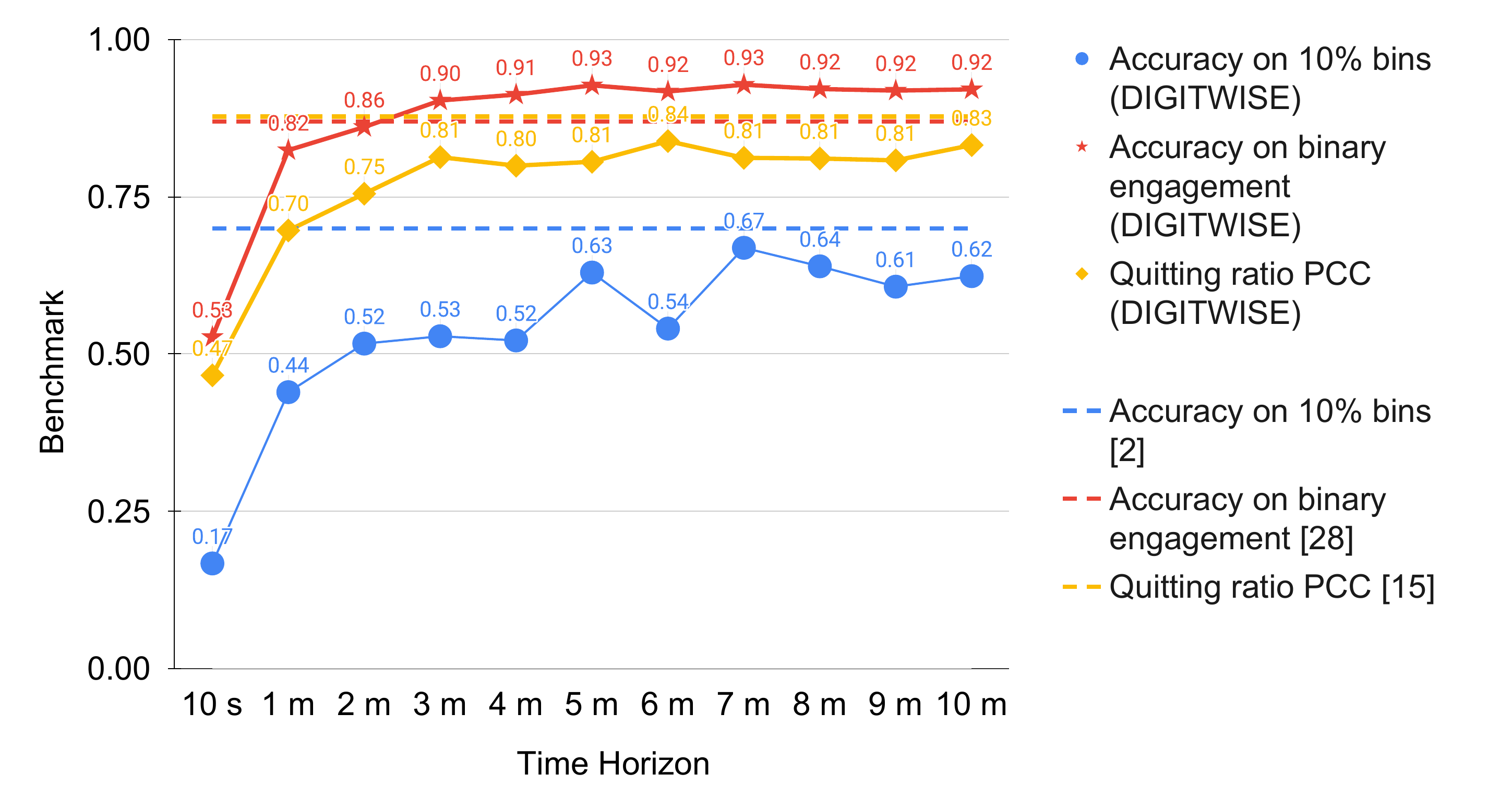}
    \caption{Comparing the performance of DIGITWISE with benchmarks~\cite{balachandran2013developing,shafiq2014network,lebreton2020predicting} over different time horizon values.}
    \label{fig:benchmarks}
\endminipage\hfill
\minipage{0.5\textwidth}
  \includegraphics[width=\linewidth]{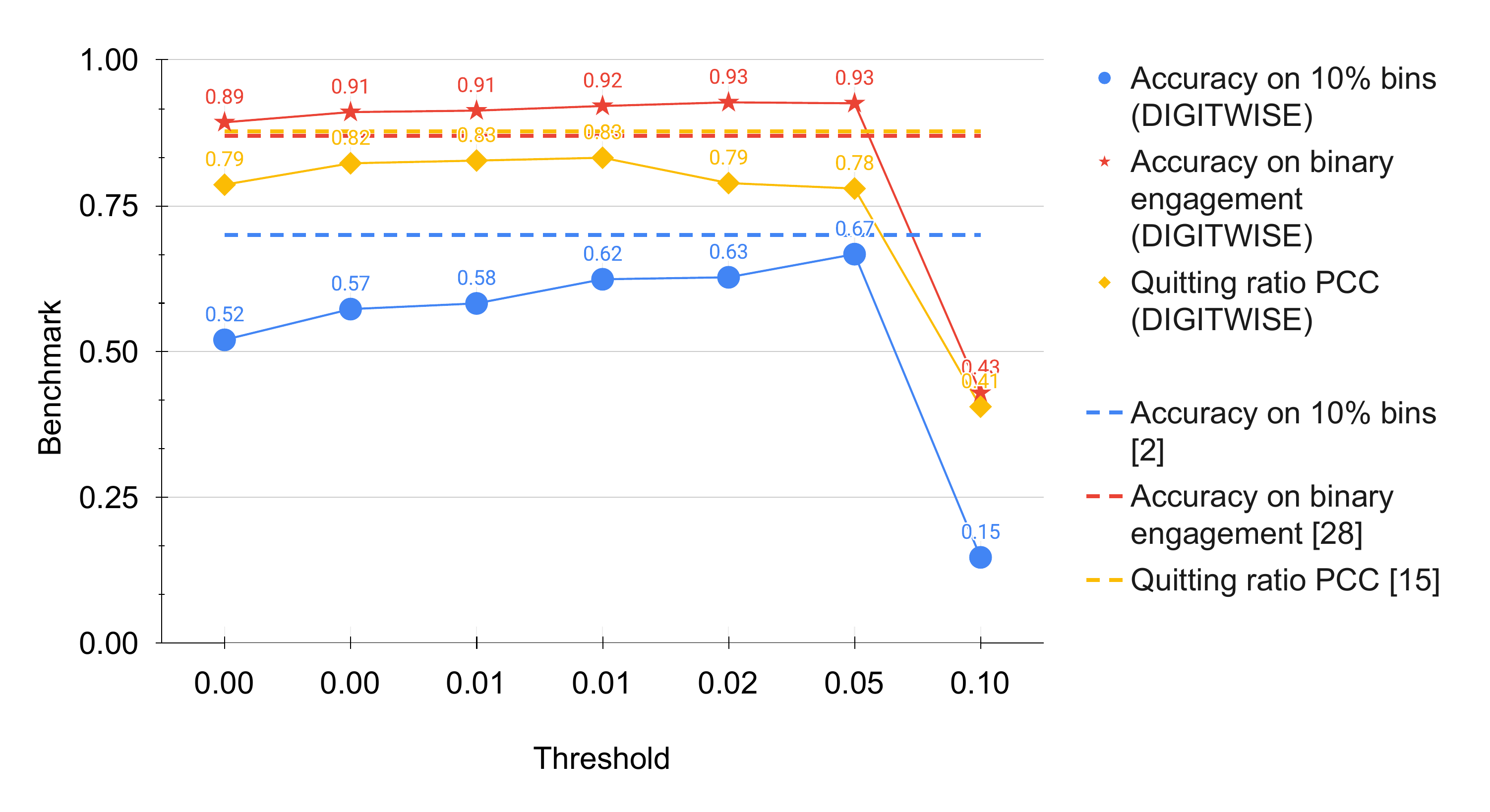}
    \caption{Comparing the performance of DIGITWISE with benchmarks~\cite{balachandran2013developing,shafiq2014network,lebreton2020predicting} over different threshold values.}
    \label{fig:benchmarks2}
\endminipage
\end{figure*}

Figures~\ref{fig:benchmarks2}-\ref{fig:benchmarks} show how DIGITWISE compares against three benchmarks:

\begin{itemize}[leftmargin=*,noitemsep,topsep=0pt]
\renewcommand{\labelitemi}{\scriptsize$\blacksquare$}
    \item In the benchmark by Balachandran~et al.~\cite{balachandran2013developing}, DIGITWISE is able to approach the target accuracy on occasion but tends towards a slightly lower level.
    \item Our data does not include information about downloaded segments. Vice versa, ~\cite{shafiq2014network} only had access to network data and could not distinguish between a video downloaded but not watched in full. 
    Therefore, a perfect comparison is not possible. To benchmark against them, we needed to identify a threshold of engagement above which to consider a session to be the equivalent of a fully downloaded. 
    Two extreme threshold values were considered: 50\%, i.e., each video watched for at least half is in the downloaded class. 
    This offered balanced classes, as we sampled our dataset uniformly over engagement. 
    However, in practice, a playback reaches 50\% way sooner than being completely downloaded, so this threshold is poorly adapted to the reality of streaming. 
    Vice versa, a threshold of 100\%, where only the videos with an engagement of almost 1 were considered for the downloaded class, is adherent to the reality of video streaming, but highly imbalanced. 
    Therefore, we chose a 70\% threshold as a middle ground among these extremes. 
    Under this hypothesis, DIGITWISE achieves a 53\% accuracy at the 10-second time horizon but improves steadily as more data is available, matching the benchmark at the 2-minute mark and reaching an accuracy of 93\%.
    \item Lebreton~et al.~\cite{lebreton2020predicting} does not provide a quantitative value for quitting "midway through". Therefore, we assume that the value is 50\% and consider sessions with an engagement higher than that as being in one class and sessions with a lower engagement to comprise the other class.
    DIGITWISE reaches 0.8 PCC with 3 minutes of data and improves slowly from there, remaining slightly lower than the benchmark of 0.87 PCC.
\end{itemize}

It should be noted that DIGITWISE was not re-trained with these new target metrics. 
Instead, we used the same regression model to obtain engagement predictions, then engineered these predictions into the same shape as the ones from the benchmark to calculate DIGITWISE performance. 
Likely, a model trained on the final prediction shape and using the same scoring metric (instead of the root mean squared error scoring used to train DIGITWISE) would achieve higher results in a specific benchmark, but the way the benchmark has been conducted was considered a more fair assessment of DIGITWISE capabilities.

\textbf{Scenario 5}.
AWS EC2 machines are utilized to generate CAdViSE sessions used in the application of the model. 
For the present scope, we consider three features, i.e., segment size, network trace, and ABR, that directly impact one or more of the engagement model's features so that even if the model does not train directly on them, it is still possible to gauge their contribution. 
This is a limitation of the current study, which ideally would include each feature of interest, but these were not present in the available data, and their inclusion is left for future work. 
The features and distribution of sessions for each combination of features are presented in Table~\ref{tab:cadvise-distribution}.
\begin{itemize}[leftmargin=*,noitemsep,topsep=0pt]
    \renewcommand{\labelitemi}{\scriptsize$\blacksquare$}
    \item ABR: The highest difference is seen with a segment size of 2 seconds and network trace Cascade-20~\cite{bentaleb2021common}(the value of $x$ in Cascade-$x$ indicates the duration of fixed bandwidth). 
    For this combination, moving from L2A~\cite{karagkioules2020online} to Bola~\cite{spiteri2020bola} increases the average engagement by 8.6\%.
    \item Network trace: The most highly represented combination is a segment size of 2 seconds, and Bola ABR. For this combination, moving from AmazonFCC~\cite{amazonfcc} to Cascade-20~\cite{bentaleb2021common} yields an average engagement increase of 7.7\%.
    \item Segment size: The most highly represented combination is an ABR L2A and a network trace AmazonFCC. For this combination, moving from a segment size of 2 seconds to 1 second increases the average engagement by 3.8\%.
\end{itemize}
Therefore, in the worst-case scenario of a video streaming platform set up with the worst parameters, modifying these parameters to those suggested by DIGITWISE would lead to an increase in engagement of more than 8.6\%.
Worthy of note is the sometimes small engagement increase when comparing two very different network traces. 
For example, moving from low bandwidth to high bandwidth increases engagement by 2.7\%, which might seem like a small improvement for such a high bandwidth difference. 
However, this is consistent with Figure~\ref{fig:model-sensitivities}, where the features impacted by the network trace do not appear in the top most important features. 
Thus, the choice of user is again proven to be significant: while better network conditions may prove critical for certain users, they are not for everyone.

\section{Conclusion}
In conclusion, this paper demonstrated how user sensitivities, while often overlooked, are an important factor in engagement considerations and how, by leveraging them,
DIGITWISE is able to significantly improve engagement modeling and predictions. 
Specifically, the Mean Average Error in engagement predictions decreased by up to 5.8\% when including user sensitivities in the model.
Furthermore, DIGITWISE is flexible and can match different state-of-the-art algorithms in their respective benchmarks without training on any specific task. 
This flexibility allows DIGITWISE to serve various applications, for example, by adding real users behaviors to simulated sessions and gauging the engagement improvements of specific design choices.
While not exhaustive, the feature set available for this experiment allowed us to suggest several design improvements that could increase user engagement by up to 8.6\%.
The DIGITWISE approach offers a valuable framework for enhancing adaptive video streaming by personalizing the user experience.
With a more comprehensive dataset and/or models capable of time-sensitive predictions, DIGITWISE's performance can be pushed even further.

\balance
\printbibliography
\end{document}